\documentclass[%
 reprint,longbibliography,
superscriptaddress,
showpacs,preprintnumbers,
 amsmath,amssymb,
pra,
floatfix]{revtex4-1}

\usepackage{graphicx}
\usepackage{dcolumn}
\usepackage{bm}

\begin{document}

\preprint{APS/123-QED}

\title{Measuring the translational and rotational velocity of particles in helical motion using structured light}
\author{Carmelo Rosales-Guzm\'an}
\email{carmelo.rosales@icfo.es}
\author{Nathaniel Hermosa}
\affiliation{ICFO-Institut de Ciencies Fotoniques, Mediterranean Technology Park, 08860 Castelldefels (Barcelona), Spain}%
\author{Aniceto Belmonte}
\affiliation{Universitat Polit\`{e}cnica de Catalunya, BarcelonaTech, Dept. of Signal Theory \& Communications, 08034 Barcelona, Spain}%
\author{Juan P. Torres$^{1,}$}
\affiliation{Universitat Polit\`{e}cnica de Catalunya, BarcelonaTech, Dept. of Signal Theory \& Communications, 08034 Barcelona, Spain}%
\date{\today}

\begin{abstract}
We measure the rotational and translational velocity components of particles moving in helical motion using the frequency shift they induced to the structured light beam illuminating them. Under Laguerre-Gaussian mode illumination, a particle with a helical motion reflects light that acquires an additional frequency shift proportional to the angular velocity of rotation in the transverse plane, on top of the usual frequency shift due to the longitudinal motion. We determined both the translational and rotational velocities of the particles by switching between two modes: by illuminating with a Gaussian beam, we can isolate the longitudinal frequency shift; and by using a Laguerre-Gaussian mode, the frequency shift due to the rotation can be determined. Our technique can be used to characterize the motility of microorganisms with a full three-dimensional movement.
\end{abstract}

\maketitle
The search for reliable methods to detect the velocity of micro- and nano-particles, and microorganisms, in a three-dimensional (3D) motion is challenging and is continually being addressed \cite{molhumreprod,jmathbiol,amerzool}. Certainly, the ability of quantifying the full velocity of a particle's movement opens new possibilities. For example, it can unveil some of the most intriguing biomechanical causes and ecological consequences of particle movement such as in helical swimming \cite{bullmathbiol2,bullmathbiol3}. Nearly all aquatic microorganisms smaller than $0.5$ mm long, exhibit helical swimming paths that are inherently three-dimensional, either in search for food, to move toward appropriate temperature or pH, or to escape from predators \cite{repprogphys,amerzool,amjphys,bullmathbiol1}. This is for instance the case of spermatozoa when traveling towards the ovum and the specific characteristics of the movement should be taken into account when creating fertilization models \cite{biobull,scirepting}
.

To characterize 3D motion, researchers usually extend two-dimensional (2D) measurement schemes which are commonly based on standard optical systems, such as video cameras and microscopes \cite{OpticaActa}. Other researchers employ numerical analysis \cite{JReprodFertil,BullMathBiol_2011}, theoretical models \cite{BiophysicalJ1972,BiophysicalJ} or digital holography with extensive numerical computations \cite{MeasSciTechnol,DiCaprio:14,PNAS}. Also, standard techniques based on the classical non-relativistic Doppler effect have been widely applied to determine velocity components perpendicular to the direction of propagation of the illuminating light beam, relying upon measurements of the longitudinal component for a large set of directions \cite{ApplOpt}. Unfortunately, many of these schemes require the use of multiple laser sources pointing at different directions, or alternatively a single laser source with fast switching of its pointing direction.

In 2011, Belmonte and Torres put forward a novel method to measure directly transverse velocity components using structured light as illumination source \cite{ol}. In their scheme, the illumination beam, which points in a fixed direction, can take on different spatial phase profiles that can be tailored to adapt to the target's motion. With their method, the determination of the transverse velocity is enormously simplified. Two recent experiments  have used structured light beams to measure the angular velocity of targets rotating perpendicularly to the direction of illumination, something that cannot be determined with commonly used Gaussian beams \cite{SciRep,science}. In \cite{SciRep}, the angular velocity of a micron-size rotating particle was measured using a Laguerre-Gauss $LG_0^{\ell}$ mode, a beam that contains an azimuthally varying phase gradient. The angular velocity of a rotating macroscopic body, on the other hand, was measured using a superposition of two $LG_0^{\ell}$ modes with opposite 
winding number $\ell$ in \cite{science}. However, both experiments restricted their attention to a 2D motion.

Here we demonstrate that this technique can also be used to detect all velocity components in a full 3D helical motion using Laguerre-Gauss ($LG_0^{\ell}$) modes, which present an azimuthally varying phase profile. This is a two-step technique wherein we first illuminate the target with a Gaussian mode to determine its translation velocity. Then we change the illumination to an $LG_0^{\ell}$ mode to obtain the velocity of rotation. For the case when the direction of translation is known, it is possible to determine the sense of rotation by simply reversing the sign of the mode index ${\ell}$ of the $LG_0^{\ell}$ mode. Conversely, if we know the sense of rotation, we can compute the direction of translation, again by reversing the sign of ${\ell}$. Even though for convenience we implement the technique in two steps, one can envision illuminating the target with the two beams simultaneously.
\begin{figure}[t]
  \centering
 \includegraphics[width=.47\textwidth]{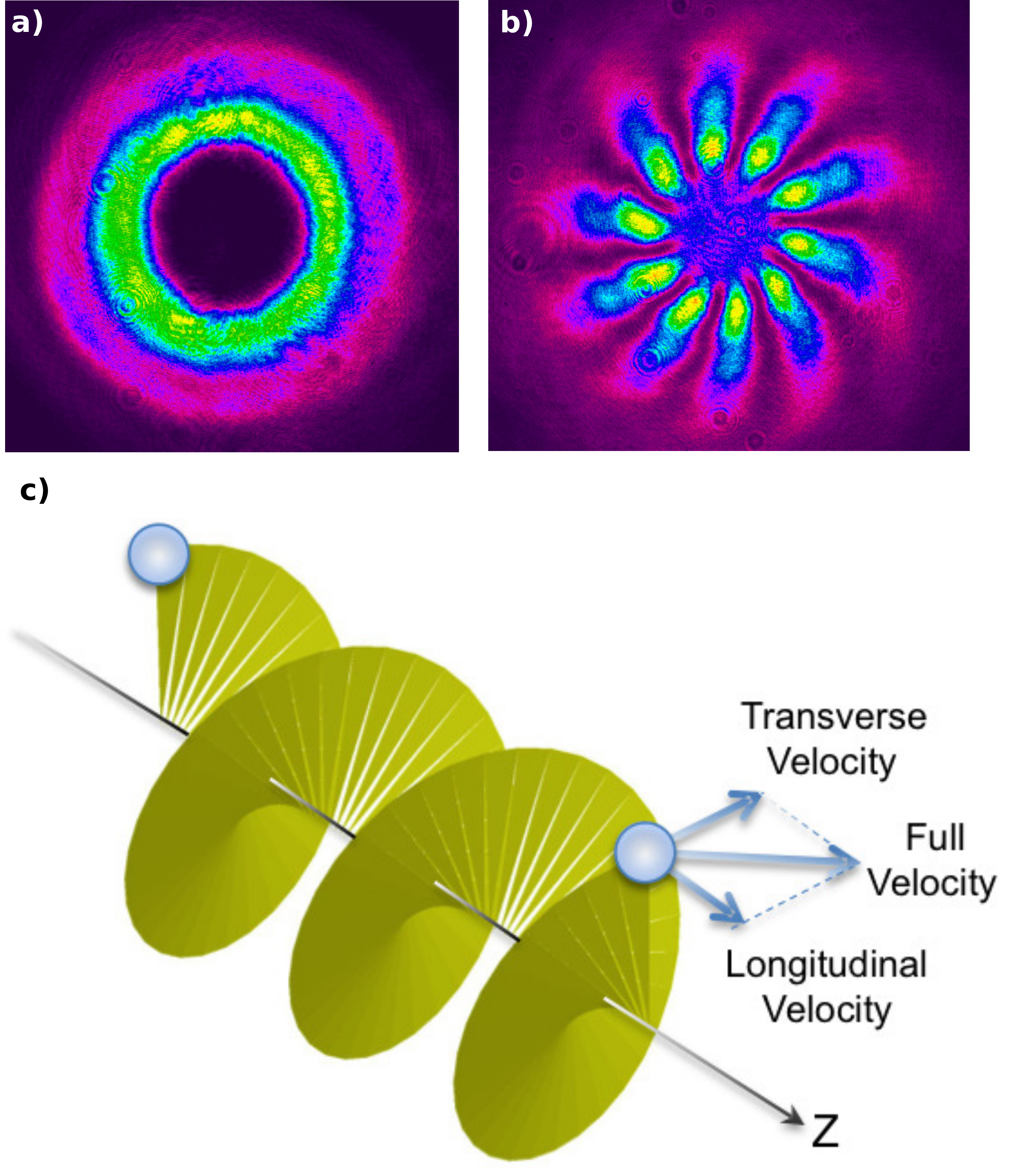}
  \caption{\small (Color online). a) Intensity profile of the $LG^{10}_0$ beam illuminating the Digital Micromirror Device (DMD). b) Interference of the $LG^{10}_0$ beam with the collimated Gaussian reference beam when all the micromirrors of the DMD reflects all incoming light. The $10$ lobes observed are due to the phase profile $\Phi=\ell \varphi$ with $\ell=10$ of the beam. c) Schematic representation of the helical trajectory followed by particles,where Z is the propagation axis of the beam.}
  \label{beam}
  \end{figure}

In the classical non-relativistic scheme, light reflected from a moving target is frequency shifted proportional to the target's velocity as $\Delta f_\parallel=2|{\bf v}|cos(\theta)/\lambda$, where $\lambda$ is the wavelength of light and $\theta$ is the angle between the velocity ${\bf v}$ of the target and the direction of propagation of the light beam. Under structured light illumination, the phase along the transverse plane is no longer constant. Therefore in the presence of a transverse velocity component, the frequency shift will have an additional term and will given by \cite{ol},

\begin{equation}
\Delta f=\Delta f_\parallel +\Delta
f_\perp=\frac{1}{2\pi}(2kv_z+\nabla_\perp\Phi\cdot{\bf v}_\perp),
\label{Doppler}
\end{equation}
where $v_z$ is the velocity of the target along the line of sight, ${\bf v}_\perp$ is the velocity in the transverse plane, $k=2\pi/\lambda$ is the wave vector and $\nabla_\perp\Phi$ is the transverse phase gradient.

Helical motion is a combination of a translation along the line of sight and a rotation in the transverse plane [see Fig. \ref{beam}(c)]. Therefore, the use of an azimuthally varying phase $\Phi=\ell\varphi$, present in an $LG$ beam [see Fig. \ref{beam}(a) and (b)], simplifies the determination of the angular velocity of rotation enormously. Here, $\varphi$ is the azimuthal angle and $\ell$ is the number of $2\pi$ phase jumps of the light beam as one goes around $\varphi$. The second term of Eq. (\ref{Doppler}) now takes the form $\nabla_\perp\Phi\cdot{\bf v}_\perp=\ell \Omega$, so that
\begin{equation}
\Delta f= \frac{1}{2\pi}(2kv_z+\ell\Omega). \label{DopplerTotal}
\end{equation}
Two aspects of this equation should be highlighted. First, if we illuminate with a Gaussian mode ($\ell=0$), as most radar Doppler systems do, $\Delta f= \Delta f_\parallel=kv_z/\pi$ and the translational velocity can be determined. Second, $\Delta f$ depends on the relative signs of $v_z$, $\ell$ and $\Omega$. It will acquire maximum value when $v_z$ and $\ell \Omega$ have the same signs while it will have a minimum value when they have opposite signs.

To generate the helical motion of particles, we employed a Digital micromirror device (DMD). A cluster of $512$ diamond-shape squares ($10\times10\,\mu$m) randomly distributed within an area with a diameter of 3 mm, was set to rotation as a solid body in a similar way to \cite{SciRep}. The DMD was attached to a translation stage (BP1M2-150 from Thorlabs, with a maximum displacement of $150$ mm), aligned perpendicular to the plane of rotation of the DMD. The axis of the illumination beam is aligned to coincide with the axis of the helical trajectory.

\begin{figure}[t]
\centering
\includegraphics[width=.47\textwidth]{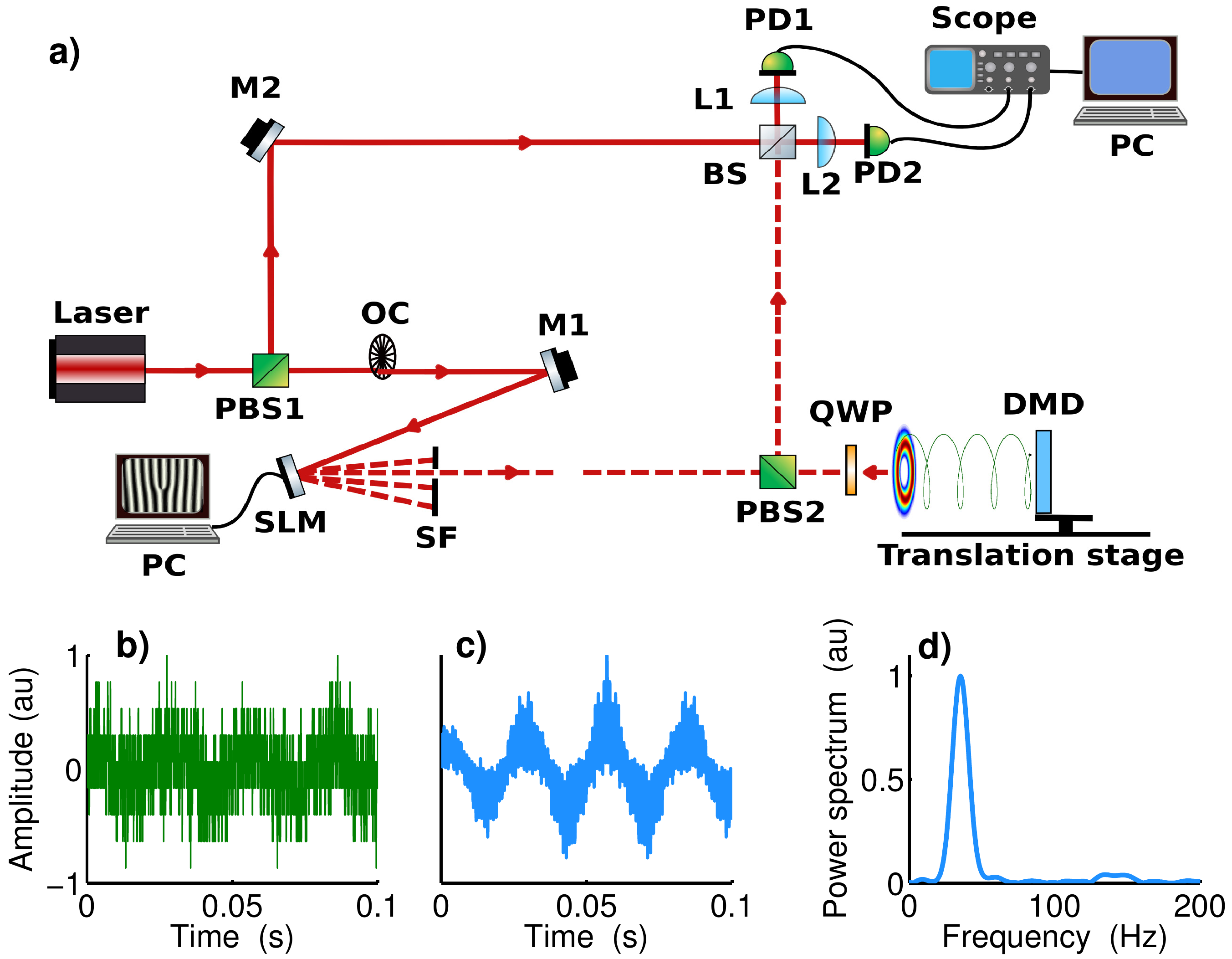}
\caption{\small  a) Experimental setup to measure all velocity components of particles with helical motion.  b) raw signal after balanced detection, c) autocorrelation function of the signal, and d) power spectral density. Here we only show the spectrum at frequencies higher than the chopping frequency. PBS: polarizing beam splitter; M: mirror; L: lens; PD: photodetector; SLM: Spatial Light Modulator; SF: spatial filter; OC: optical chopper; QWP: Quarter-Wave  Plate; DMD: Digital Micromirror Device. {\it See text for details.}} \label{setup}
\end{figure}

We extract the Doppler frequency shift imparted by the moving particles via an interferometric technique using the modified Mach-Zehnder interferometer shown in Fig. \ref{setup}(a). A $15$ mW continuous wave He-Ne laser (Melles-Griot, $\lambda=632.8$ nm) is spatially cleaned and expanded to a diameter of $5$ mm. This beam is split into two (the {\em signal} and {\em reference} beams) using a polarizing beam splitter (PBS1). A mirror (M1) redirects the signal beam to a Spatial Light Modulator (SLM, LCOS-SLM X10468-02 from Hammamatsu) that imprints the beam with the structured phase, as shown in Fig. \ref{beam}. The first diffracted order of the fork-like hologram encoded into the SLM is used to illuminate the target while the rest of the diffracted orders are spatially filtered (SF). A second polarizing beam splitter (PBS2) in combination with a quarter-wave plate (QWP) collects light reflected from the target back into the interferometer. These reflections are afterwards interfered with the reference signal 
using a beam splitter (BS).

A balanced detection is implemented with two photodetectors (PD1 and PD2). These are connected to an oscilloscope (TDS2012 from Tektronix). An optical chopper (OC) placed in the path of the signal beam shifts our detected frequency from Hz to kHz, so that we can eliminate low frequency noise. Figure \ref{setup}(b) shows a typical signal resulting from the difference of the signals detected from output ports PD1 and PD2. An autocorrelation process allows us to improve the signal-to-noise ratio significantly by finding periodic patterns obscured by noise [Fig. \ref{setup}(c)]. The resulting signal is Fourier transformed to find the fundamental frequency content. Hamming windowing and zero padding are also applied to smooth the Fourier spectrum, shown in Fig. \ref{setup}(d).

\begin{figure}[t]
  \centering
 \includegraphics[width=.48\textwidth]{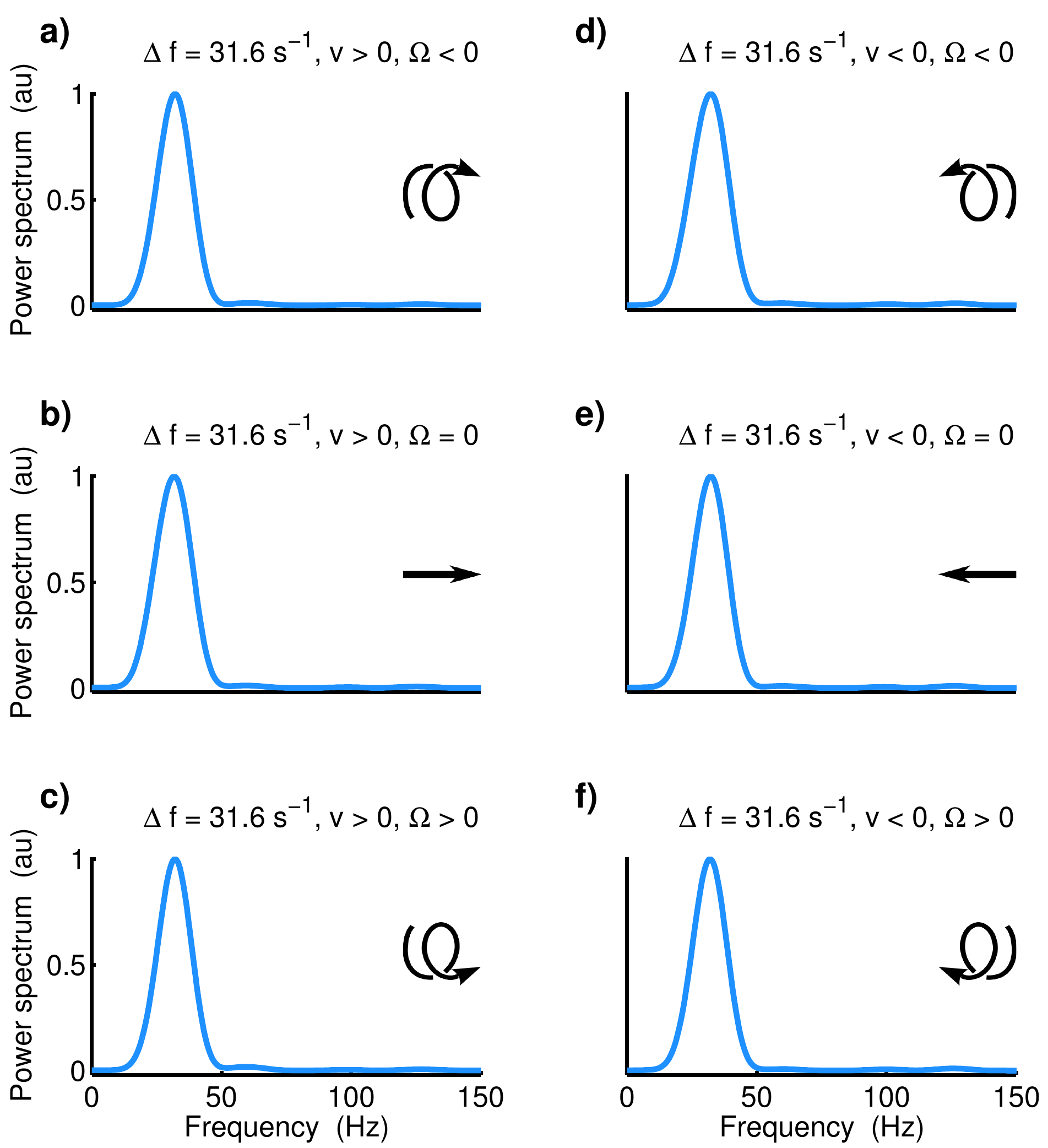}
   \caption{\small Frequency shift measured when particles are illuminated with a Gaussian mode ($\ell=0$) for any direction of
   translation($v_z > 0$ and $v_z < 0$), and any sense of rotation ($\Omega > 0$ and ($\Omega < 0$). For the sake of comparison, the
   case with $\Omega=0$ is also shown.}
  \label{translation}
  \end{figure}

Due to the interferometric nature of the experimental set up considered here, we always measure $|\Delta f|$. However, since the frequency shift given by  Eq. (\ref{DopplerTotal}) is dependent on the relative signs of $v_z$, $\ell$, and $\Omega$, we can determine the absolute magnitudes of $|\Delta f_{\parallel}|$ and $|\Delta f_{\perp}|$, and consequently $|v_z|$, $|\Omega|$ and the relative sign between $v_z$ and $\Omega$.  The frequency shifts always fulfill $|\Delta f| > |\Delta f_{\parallel}|$ for $v_z$ and $\ell\Omega$ showing the same sign, while $|\Delta f| < |\Delta f_{\parallel}|$ for $v_z$ and $\ell\Omega$ showing opposite signs. Therefore, if $|\Delta f|$ is larger for $\ell>0$ than for $\ell<0$, $v_z$ and $\Omega$ have the same sign, while if is smaller, $v_z$ and $\Omega$ have opposite signs. The important point is that the sign of $\ell$ is a free parameter imposed on the illumination beam chosen in the experiment. The experimental procedure to measure $|v_z|$, $|\Omega|$ and the sign of $v_z \cdot\Omega$ is thus to choose a beam with $\ell=0$ to obtain $|\Delta f_{\parallel}|$, use afterward a $LG$ beam with the sign of the index $\ell$ so that it maximizes $|\Delta f|$, and obtain $|\Delta f_{\perp}|=|\Delta f|-|\Delta f_{\parallel}|$. Furthermore, if $\ell>0$, we have $v_z \cdot \Omega>0$, and $v_z \cdot \Omega<0$ otherwise. Previous knowledge of the sign of $v_z$, for instance knowing that the particle or microorganism under study advances in a fluid stream, allows to determine the sign of $\Omega$.

\begin{figure}[t]
 \centering
\includegraphics[width=.48\textwidth]{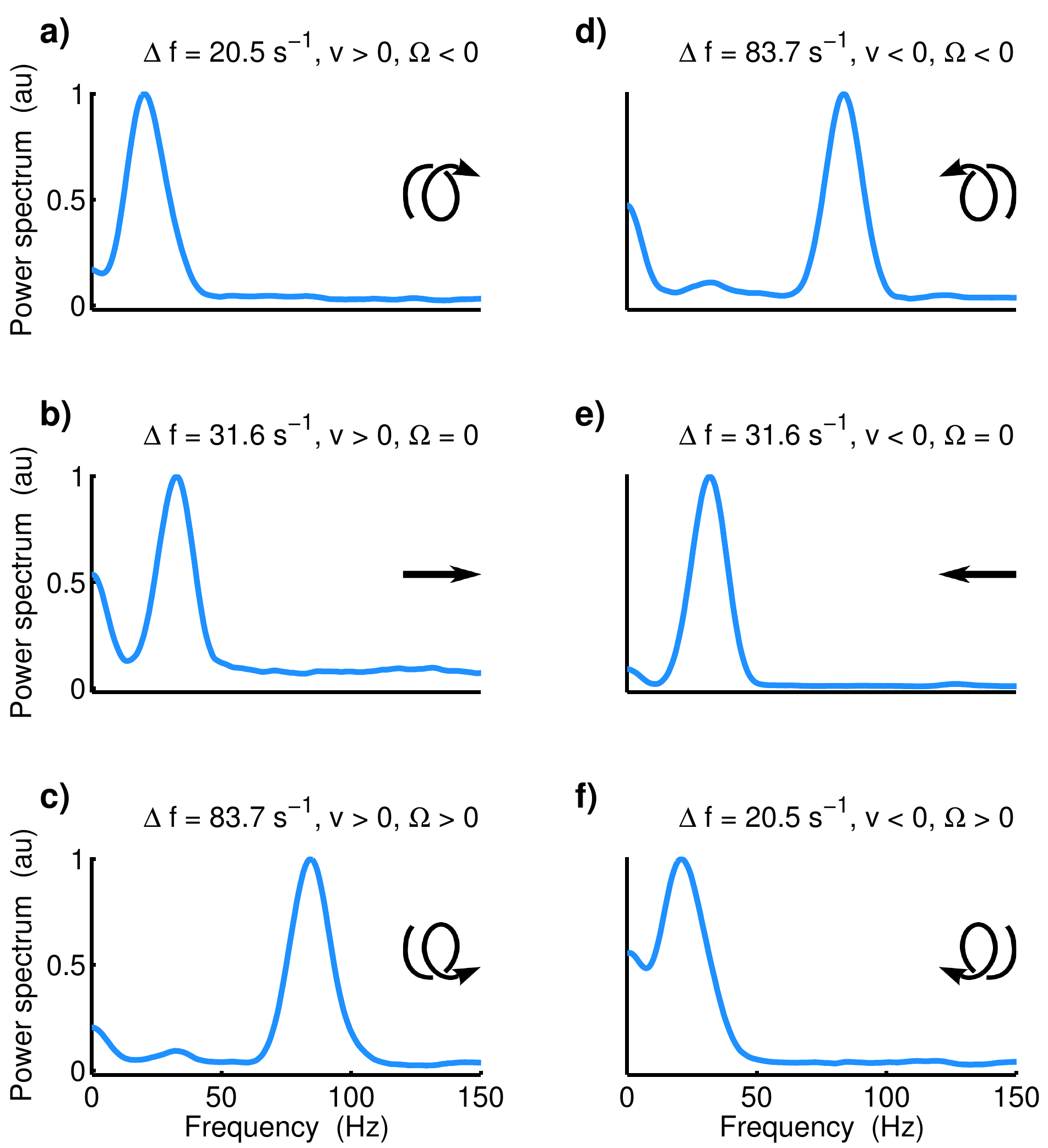}
 \caption{Frequency shift measured when particles are illuminated with a Laguerre-Gauss mode ($\ell=-10$) for any direction of translation ($v_z > 0$ and $v_z < 0$) and any sense of rotation ($\Omega > 0$ and ($\Omega < 0$).}
 \label{m=-10}
 \end{figure}
\begin{figure}[t]
 \centering
\includegraphics[width=.48\textwidth]{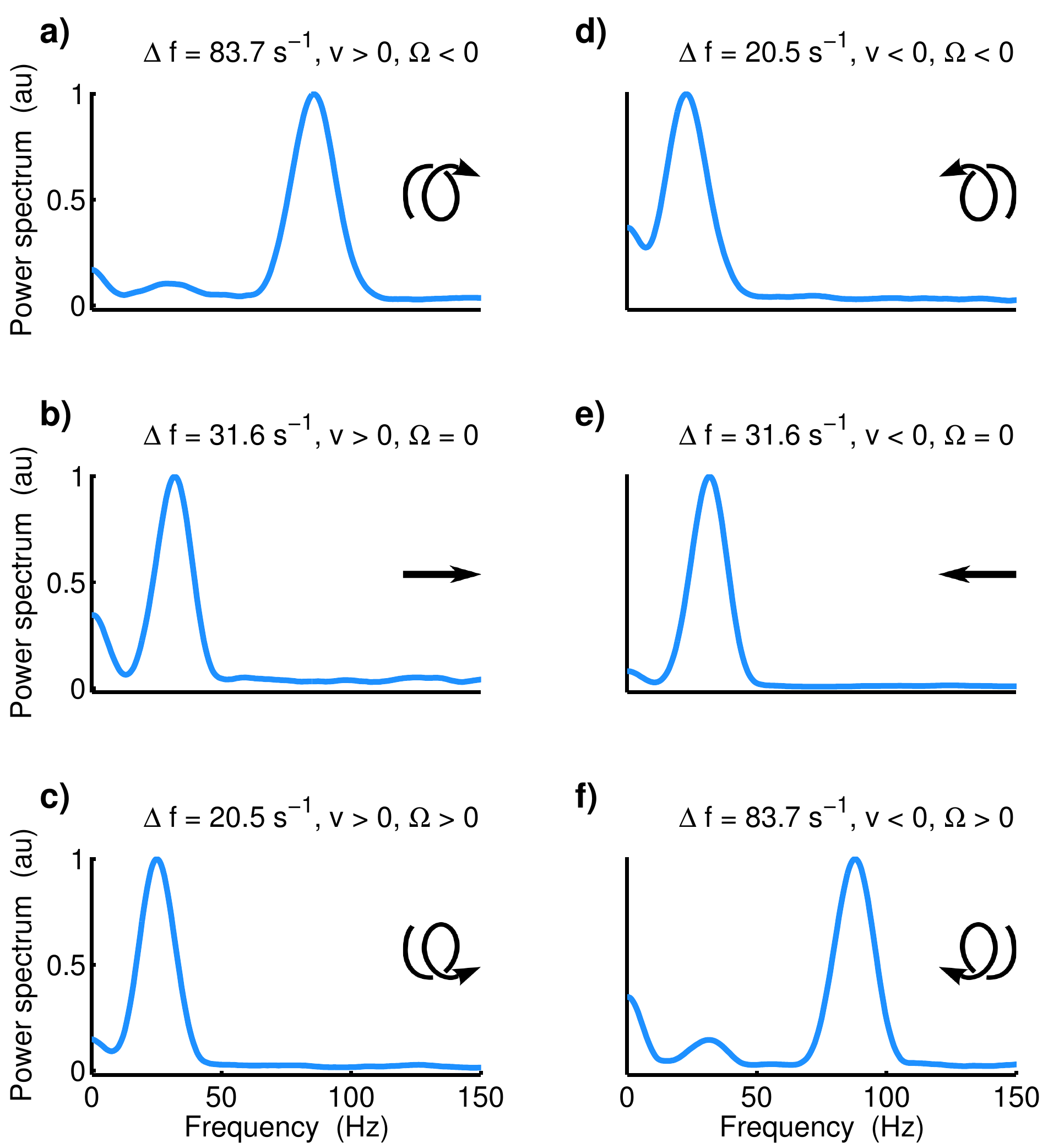}
 \caption{Frequency shift measured when particles are illuminated with a Laguerre-Gauss mode ($\ell=10$) for any direction of translation ($v_z > 0$ and $v_z < 0$) and any sense of rotation ($\Omega > 0$ and ($\Omega < 0$).}
 \label{m=10}
 \end{figure}

In our experiments, we first illuminate the helically moving particles with a Gaussian beam, so that $\Delta f=\Delta f_{\parallel}$, since the particles see a constant phase along the transverse plane. Figure \ref{translation} shows the experimental results. We do all possible combinations: $v_z>0$ (forward, moving away from the illuminating source) with negative [Fig. \ref{translation}(a)] and positive [Fig. \ref{translation}(c)] rotations, as well as $v_z<0$ (backwards, moving towards the illuminating source) with negative [Fig. \ref{translation}(d)] and positive [Fig. \ref{translation}(f)] rotation. Positive (negative) rotation refers to anticlockwise (clockwise) rotation for an observer looking towards the illumination source. For the sake of comparison, we also show the spectra when particles translate without any rotation for both $v_z>0$ and $v_z<0$ [Figs. \ref{translation}(b) and \ref{translation}(e), respectively]. The frequency shift measured, $|\Delta f_{\parallel}|=$ 31.6 s$^{-1}$, is the same 
in all cases, regardless of the relative signs of $v_z$ and $\ell$. From this information, computing the translational velocity is straight forward from Eq. (\ref{DopplerTotal}). The velocity we obtained is $|v_z|= \pm 10\,\mu$m/s, exactly the same as the velocity we programmed our rail to move. This leaves only the rotational velocity unknown.

Figures \ref{m=-10} and \ref{m=10} show experimental results when the moving particles are illuminated by a $LG_0^{\ell}$ mode with $\ell= \pm 10$. For $\ell=-10$ (Fig. \ref{m=-10}) and $v_z > 0$, $\Delta f$ is larger if the particles rotate with $\Omega<0$ [Figs. \ref{m=-10}(a)], while it is smaller if the particles rotate in the opposite direction ($\Omega>0$) [Fig. \ref{m=-10}(c)]. The frequency shifts measured are $\Delta f=$ 83.7 s$^{-1}$ and $\Delta f= $20.5 s$^{-1}$ for $\Omega<0$ and $\Omega>0$, respectively. By accounting these and using the procedure described above, we can compute $\Delta f_{\perp}$ due to the rotation of the particles according to Eq. (\ref{DopplerTotal}). In all cases, $|\Delta f_\perp|=$ 52.1 s$^{-1}$, yielding an angular velocity of $|\Omega|=$32.7 s$^{-1}$. Conversely, when the particles move backwards ($v_z<0$), $\Delta f$ is larger for $\Omega>0$ [Fig. \ref{m=-10}(f)], while it is smaller for $\Omega<0$ [Fig. \ref{m=-10}(d)]. On the contrary, if we illuminate with $LG_0^{10}
$, we observe a larger $\Delta f$ when either $v_z > 0$ and $\Omega > 0$ [Fig. \ref{m=10}(c)] or $v_z<0$ and $\Omega<0$ [Fig. \ref{m=10}(d)]. We also show the spectra when the particles move forward [Fig. \ref{m=-10}(b) and \ref{m=10}(b) ] or backward [Figs. \ref{m=-10}(e) and \ref{m=10}(e)] without rotation ($\Omega=0$).

The translational and rotational velocities used in our experiment are within the range of velocities typical to that of biological specimens. Dinoflagellates, for example, have typical free-swimming velocities of around $20$-$100 \mu$m/s \cite{MeasSciTechnol}. Human sperms have helical rotation speeds of approximately $3$ to $20$ rotations/s ($\sim18$ to $120$ s$^{-1}$) and linear speeds of approximately $20$ to $100\mu$m/s \cite{DiCaprio:14,PNAS}.  And lastly, it has been reported that blood in the retina flows in the range of some tens of mm/s on average \cite{JBiomedOpt12}.

In conclusion, we have measured all components of the velocity of particles moving in a helical motion by using the frequency shift generated by reflecting particles moving in a structured light beam, i.e., a beam with an engineered transverse phase profile. This technique only requires choosing the appropriate spatial mode of the illumination beam, which depends on the characteristics of the particular type of movement under investigation. First, we determine the longitudinal velocity component by illuminating with a Gaussian mode and afterwards, we use an $LG_0^{\ell}$ mode to obtain the angular velocity of the rotational motion.

Even though we have characterized here the movement of a particle with helical motion, the technique can be easily generalized to characterize the movement of particles with different types of motion by choosing other types of illuminating beams with different transverse phase profiles.

\begin{acknowledgments}
This work was supported by the Government of Spain (project FIS2010-14831 and program SEVERO OCHOA), and the Fundacio Privada Cellex Barcelona.
\end{acknowledgments}

\end{document}